# Efficient Dynamic Searchable Encryption with Forward Privacy

Mohammad Etemad  Alptekin Küpçü  Charalampos Papamanthou  David Evans
University of Virginia  Koç University  University of Maryland  University of Virginia
etemad@virginia.edu  akupcu@ku.edu.tr  cpap@umd.edu  evans@virginia.edu

**Abstract**

Searchable symmetric encryption (SSE) enables a client to perform searches over its outsourced encrypted files while preserving privacy of the files and queries. Dynamic schemes, where files can be added or removed, leak more information than static schemes. For dynamic schemes, *forward privacy* requires that a newly added file cannot be linked to previous searches. We present a new dynamic SSE scheme that achieves forward privacy by replacing the keys revealed to the server on each search. Our scheme is efficient and parallelizable and outperforms the best previous schemes providing forward privacy, and achieves competitive performance with dynamic schemes *without* forward privacy. We provide a full security proof in the random oracle model. In our experiments on the Wikipedia archive of about four million pages, the server takes one second to perform a search with 100,000 results.

## 1 Introduction

Searchable symmetric encryption (SSE) enables a data owner to outsource private data to an untrusted server, while selectively retrieving data elements matching a query without revealing either the data contents or the search keywords to the server. Although asymmetric searchable encryption schemes have been proposed [3, 4], we focus on schemes providing high efficiency using symmetric encryption.

A naïve solution is to outsource encrypted files to the server, and store locally an inverted index that associates each keyword to the set of identifiers of files sharing the keyword. The client finds the list of files matching a desired query, and retrieves those encrypted files from the server. This requires the client to store a large index and perform all required search and update computations locally. To reduce the client storage and computation, the index is encrypted and outsourced to the server. To operate on the index, the server requires a *token* generated by the client using its secret key. As the tokens are deterministic, the server learns if multiple searches involve the same keyword. This is known as *search pattern* leakage [12, 22].

Efficient symmetric searchable encryption schemes [12, 19, 9, 8, 25, 28, 6] achieve optimal asymptotic search cost $O(d)$, where $d$ is the number of files in the result. However, this is achieved at the cost of leaking the list of files sharing the keyword. This is known as the *access pattern* leakage [11].

These leakages can be exploited by the adversarial servers to compromise privacy of the data and queries. Several recent papers have shown how such leakage can be exploited to learn sensitive information about the queries or file contents [17, 22, 1, 7, 24, 31]. These attacks demonstrate that the privacy provided by traditional SSE schemes does not satisfy expectations in practice. On the other hand, data owners want to fully utilize cloud data services, and rest assured that the privacy of their data (e.g., emails, business data) is preserved.

Stefanov *et al*. [28] asserted that a secure SSE scheme must satisfy both forward and backward privacy. These two properties capture common expectations for file addition and deletion, where a user expects that the privacy of newly added files in the presence of previous queries (*forward privacy*) and the privacy of deleted files once they are deleted (*backward privacy*) should be preserved. No efficient scheme is currently known that provides backward privacy, and we do not consider that in this paper. Our focus is on achieving forward privacy with an efficient SSE scheme supporting dynamic updates.

***Forward privacy*** is a strong property that states the server cannot realize whether or not a newly added file contains any of the keywords used in previous searches. SSE schemes achieving forward privacy make adaptive attacks less effective [31]. The first scheme supporting forward privacy is given by Stefanov *et al*. [28]. The search and update costs of this scheme are $O(d \log^3 N)$ and $O(r \log^2 N)$, respectively, where $r$ is the number of unique keywords in the file, $d$ is the number of files in the result, and $N$ is the number of all (keyword, file ID) mappings. It requires $O(\sqrt{N})$ client storage. Bost's Sophos [6] supports forward privacy by employing trapdoor permutations. Though the search and update costs are asymptotically optimal ($O(d)$ and $O(r)$, respectively), the client needs to run $O(r)$ asymmetric cryptographic operations on a file insertion.

*Parallelism* is an important efficiency factor that is currently supported by some schemes (e.g., [18, 28]). It requires the encrypted index be organized in a way that server can access the required parts directly. The provider can distribute the work on available servers to improve the performance [27, 2]. Our scheme supports parallelism by design.

**Contributions.** Our main contribution is designing the first asymptotically-optimal parallelizable dynamic SSE scheme that provides forward privacy. Our scheme outperforms the existing schemes providing forward privacy [28, 6], and is competitive with the most efficient dynamic SSE schemes *without* forward privacy [19, 18] while providing stronger security. In particular, our scheme:

– provides *forward privacy*: On each search, a key is revealed to enable the server operate on the encrypted index. We "revoke" this key, remove the index entries accessed, and re-insert them encrypted under a fresh key. The server never holds a valid key after a search, and hence, cannot decrypt any part of the updated index to see if a file added later contains a keyword used in a previous search.

– is *asymptotically optimal*: The update cost is $O(r)$ and the search cost is $O(d)$.

– is *parallelizable*: The server's index is a dictionary of encrypted (key, value) pairs where the keys are generated as outputs of a hash function. Each hash function evaluation is independent, that allows the load to be distributed over $p$ processors to achieve asymptotically optimal search ($O(d/p)$) and update ($O(r/p)$) cost.

– is *efficient*: The server only evaluates $O(d)$ hash functions for a search. For a file insertion, the server only inserts the $O(r)$ values given inside the token into the index. Unlike Sophos [6], no asymmetric operations are needed.

– is easily convertible to a scheme in the *standard model* by replacing the hash functions with pseudorandom functions (Section 7). This increases the search token sizes, while reducing the server load.

**Approach overview.** The server stores an encrypted index that associates each keyword with a set of file identifiers representing the files that contain that keyword. The entries associated with a keyword $w$ are encrypted with a key $K_w$ that depends on both $w$ and the number of times $w$ has been searched for. $K_w$ is derived using a pseudorandom function from a master key that is stored by the client. So, the client must store the number of times each keyword has been searched for so far. We do this in a dictionary called SearchCnt. To search for a keyword, the client generates and reveals the key $K_w$ to enable the server to operate on the appropriate entries in the encrypted index, find the identifiers of files sharing the keyword, and return the corresponding (encrypted) files.

Incrementing the number of times $w$ has been searched for on each search leads to a new key $K_w$ be generated for $w$ and invalidates the previous key revealed to the server. This ensures the freshness of $K_w$ on each search. Therefore, if $w$ appears in a new file being added, the corresponding entry will be encrypted under a fresh key and the server cannot link it to the previous searches and realize that the new file contains $w$. This provides the essential *forward privacy* property. On the downside, this requires another round of interaction (at the end of search), to encrypt the accessed index entries with the new key and upload them back to the server. (We ask the server to remove the accessed entries from the index during a search. The server can keep the deleted entries, but they have already been leaked and contain no new information.) Note that this does not increase the asymptotic search cost. Besides, we can eliminate the extra round using piggybacking (as in TWORAM [14]) and upload the current updated entries together with the next search token. The whole process ensures that no entry of the outsourced index is encrypted under a revealed key. Further, the revealed keys will never be used again.

Another requirement for forward privacy is that the identifiers of all files containing a keyword cannot be stored in an easily linkable fashion (e.g., in a set or a file). Otherwise, adding a new file would trivially reveal which of the previously searched for keywords are contained in this new file. It may further leak information about other files the new file shares keywords with. This requires the identifiers of all files containing each keyword $w$ to be stored at random locations in the index that are also determined by $K_w$.

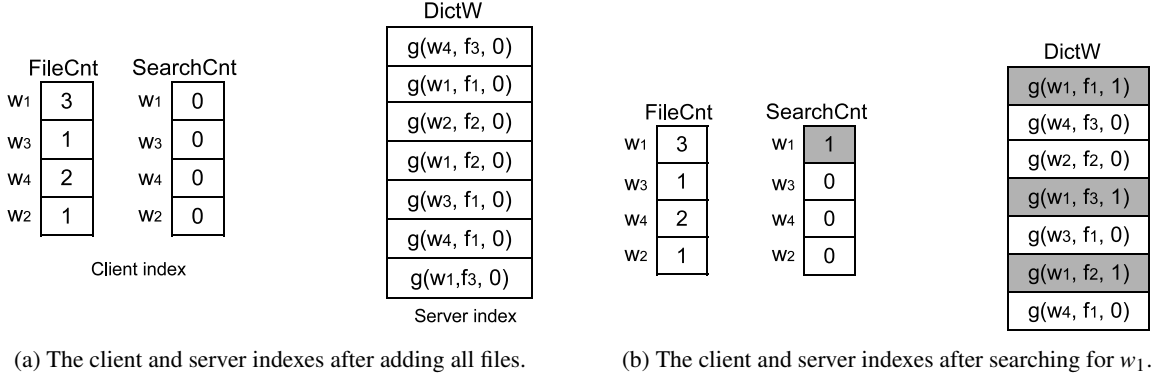

(a) The client and server indexes after adding all files.  (b) The client and server indexes after searching for $w_1$.

Figure 1: The example scenario with three files and four keywords.

This solution immediately enables parallelism for efficiency. A sequence number is assigned to each file ID among the set of files containing a given keyword. These sequence numbers are used to generate the addresses at which the respective encrypted file IDs will be stored. Therefore, given the total number of file IDs, the provider can divide it by the number of available servers and ask each one to extract a subset of file IDs to be returned. The client again needs to store the total number of file IDs sharing a keyword in a dictionary named FileCnt (detailed in Section 3).

Our scheme can be extended to support deletion, as discussed in Section 6. This improves efficiency over keeping deleted files in the index and filtering out deleted responses, but does not provide backward privacy (which remains an important, but elusive goal). An important consequence of removing the index entries on each deletion is that our scheme keeps the index up-to-date, and there is no need to do periodic rebuilds to remove the deleted entries and cleanup the index, as in other schemes [8].

Note that the size of both SearchCnt and FileCnt is $O(m)$ for a total number of $m$ keywords. Hence, it is reasonable to store both of them even at the client. In our construction, we assume they are stored by the client, and later show how to outsource them as well (Section 3.3).

**Example.** We give an example to better illustrate the client and server indexes and how the protocols operate on them. Assume there are three files and four keywords: $f_1$ contains $w_1, w_3, w_4$, $f_2$ contains $w_1, w_2$, and $f_3$ contains $w_1, w_4$. Let $g(w, f, i)$ denote the masked version of the ID of a file $f$ after the $j^{th}$ search for $w$, to be stored in DictW. Hence, $g(w_1, f_1, 0)$, $g(w_3, f_1, 0)$, and $g(w_4, f_1, 0)$ are added into DictW for $f_1$. Figure 1a shows the client and server indexes when all files are processed.

Now, the client searches for $w_1$. It sends the server the key $k_1$ and the number of files $w_1$ appears in, 3. The server locates the given number of DictW entries, decrypts the contents, and finds out that $f_1$, $f_2$ and $f_3$ are the target files. It deletes the accessed entries from DictW and sends the files to the client. The client increments the number of times $w_1$ is searched for as: SearchCnt$[w_1]$++. Then, it re-encrypts the received pairs with a fresh key generated using the updated SearchCnt$[w_1]$, and sends them back to the server. The server stores them in their new locations in DictW. The client and server storage after this operation are shown in Figure 1b.

## 2 Background

This section introduces our notation and provides a formal model for SSE and its security definitions.

### 2.1 Preliminaries

**Notation.** We use $x \leftarrow X$ to show $x$ is sampled uniformly from the set $X$, $|X|$ to represent the number of elements in $X$, and $||$ to show concatenation. $\lambda$ is the security parameter, and PPT stands for probabilistic polynomial time. A function $\nu(k) : Z^+ \to [0, 1]$ is *negligible* if $\forall$ *positive polynomials* $p$, $\exists$ *constant* $c$ such that $\forall\ k > c$, $\nu(k) < 1/p(k)$.

**File collection.** The client owns $n$ files $F = \{f_1, f_2, ..., f_n\}$, each with an identifier $id(f_j)$. The files are encrypted using a CPA-secure symmetric encryption scheme, making the encrypted files $C = \{c_1, c_2, ..., c_n\}$, where $c_j = \text{Enc}_K(f_j)$. In a dynamic setting, we represent the set of outsourced files at time $t$ as $F_t$. The set of all $m$ unique keywords in all files in $F$ is represented as $W = \{w_1, w_2, ..., w_m\}$. $N$ is the number of all existing mappings from $W$ to $\{id(f_j)\}_{j=1}^n$.

**Interactive protocols.** We describe our scheme as a set of interactive protocols demonstrating the interaction between the client and the server to perform a functionality: $(Out_{client})(Out_{server}) \leftarrow \texttt{Protocol}(In_{client})(In_{server})$.

**Symmetric-key Encryption.** A symmetric-key encryption scheme $\texttt{SKE} = (\texttt{Gen}, \texttt{Enc}, \texttt{Dec})$ consists of three PPT algorithms. $\texttt{Gen}$ takes the security parameter as input and generates a key. $\texttt{Enc}$ receives the key and a message $m$ as input, and encrypts $m$ to the respective ciphertext $c$. $\texttt{Dec}$ takes as input a key and a ciphertext $c$, and retrieves the message $m$. SKE is required to be *CPA-secure*. Refer to Katz and Lindell [20] for formal definitions.

**Pseudorandom function (PRF).** Let $\texttt{GenPRF}(1^\lambda) \in \{0,1\}^\lambda$ be a key generation function and $G: \{0,1\}^\lambda \times \{0,1\}^{l'} \to \{0,1\}^l$ be a family of pseudorandom functions mapping $l'$-bit strings to $l$-bit strings. Define $G_s(x) = G(s,x)$. G is a PRF family if $\forall$ PPT distinguishers $D$, $\exists$ a negligible function $\nu(.)$ such that: $|\Pr[s \leftarrow \texttt{GenPRF}(1^\lambda) : D^{G_s(.)}(1^\lambda) = 1] - \Pr[D^{g(.)}(1^\lambda) = 1]| \leq \nu(\lambda)$, where $g(.)$ is a truly random function [13, 20].

**Hash function.** Members of a hash function family $h : \mathcal{K} \times \mathcal{M} \to \mathcal{C}$ are identified by a $K \in \mathcal{K}$ as $h(k,.)$. A hash function family is *collision resistant* if for all PPT adversaries $\mathcal{A}$, there exists a negligible function $\nu(.)$ such that: $\Pr[K \leftarrow \mathcal{K}; (x,x') \leftarrow \mathcal{A}(h,K) : (x' \neq x) \wedge (h(K,x) = h(K,x'))] \leq \nu(\lambda)$.

## 2.2 Model

We employ a two-party model including a *client* (*data owner*) and a *server*. The client generates an inverted index mapping each keyword to the set of identifiers of files containing it, encrypts the index, and uploads it along with the encrypted files to the server. The server stores the encrypted index and files, and responds to the client's queries. The server is relied on to provide highly-available and reliable storage, but not with any confidential client data. We assume a single-client model.

Since the index and files are encrypted, the server does not learn the search keyword or the contents of files. However, by running client's queries and commands, some information leaks to the server over the time. We define the precise leakage of our scheme in Section 2.3.

**Adversarial model.** We assume an honest-but-curious server and achieve forward privacy in an efficient and parallelizable manner. In fact, since the client knows the response size, our scheme could be adapted to malicious settings with minimal overhead by adding message authentication codes to entries as proposed by Kurosawa and Ohtaki [21]. For simplicity, we do not consider those extensions here or in our security proofs.

**Definition 2.1** *A dynamic SSE scheme consists of the following PPT protocols:*

- $(sk)() \leftarrow \texttt{Gen}(1^\lambda)(1^\lambda)$: *The client starts the protocol to generate a secret key sk given the security parameter $\lambda$.*

- $(\mathcal{I}_c)(\mathcal{I}_s, C) \leftarrow \texttt{Build}(sk, F)()$: *The client starts this protocol to outsource a collection of files F given the secret key sk. It generates the index $\mathcal{I}_c$. Also, the server outputs the index $\mathcal{I}_s$ and the encrypted files C.*

- $(\mathcal{I}'_c)(\mathcal{I}'_s, C') \leftarrow \texttt{Add}(sk, f, \mathcal{I}_c)(\mathcal{I}_s, C)$: *The client starts this protocol to outsource a new file f given the secret key sk, and her current index $\mathcal{I}_c$. It updates the index to $\mathcal{I}'_c$. Similarly, the server takes his current index $\mathcal{I}_s$ and the encrypted file collection C as input, and outputs the updated index $\mathcal{I}'_s$ and the updated file collection $C'$.*

- $(\mathcal{I}'_c, F_{w,t})(\mathcal{I}'_s) \leftarrow \texttt{Search}(sk, \mathcal{I}_c, w)(\mathcal{I}_s, C)$: *This is a protocol to find and return the encrypted files containing a keyword w. The client takes as input the secret key sk, her index $\mathcal{I}_c$, and the keyword w. It updates the local index to $\mathcal{I}'_c$ and outputs the existing files $F_{w,t}$ containing w. The server receives his index $\mathcal{I}_s$ and the encrypted file collection C, and updates his index to $\mathcal{I}'_s$.*

The $\texttt{Build}$ and $\texttt{Add}$ protocols are non-interactive: The client prepares the commands and sends them to the server for execution. But $\texttt{Search}$ is an interactive protocol: The client and server work interactively to perform the computation on the encrypted indexes. Since we presented the scheme as a set of interactive protocols, there is no visible use of tokens. However, the client prepares and sends tokens during the execution of $\texttt{Add}$ and $\texttt{Search}$ to the server.

## 2.3 Security Definitions

First, we define the leakage functions that are used inside the definitions. An SSE scheme is secure if it reveals no information, even when dynamic operations are executed. Naveed *et al.* [23] observed that this level of security requires the whole outsourced index and files be transferred on each operation. Existing SSE schemes leak some information for efficiency. Moreover, dynamic operations reveal extra information such as the relation between the entries on the encrypted index being accessed and the file under operation. In the following definitions, we assume the client has issued $t$ search queries $Q = \{q_1, q_2, ..., q_t\}$ up to time $t$.

**Definition 2.2 (Search pattern)** *A search pattern is a vector,* SP*, that shows which keyword each query $q_j$ corresponds to.* $\text{SP}[j] = w_i$ *means that $w_i$ was queried at time $j$.*

Even though we re-key after each search, the search pattern still leaks since the server knows that the newly re-keyed items correspond to the completed search. The re-keying is used for forward privacy, not for hiding the search pattern. (The keywords in queries in this definition are encrypted; i.e., the server does not see the real keywords.)

**Definition 2.3 (Temporal access pattern)** *The temporal access pattern of a keyword $w$ at time $t$ is defined as the set of existing files at time $t$ sharing $w$: $F_{w,t} = \{id(f) : w \in f \land f \in F_t\}$.*

**Definition 2.4 (Access pattern)** *The access pattern in an SSE scheme is the union of all temporal access patterns of all keywords searched for so far [12].*

Observe that once a search is done for $w$, the server learns the set of files in which $w$ appears (the temporal access pattern). Later on, even when one of these files is deleted, the server knows that the file contains this keyword $w$ (though not necessarily knowing what $w$ actually is), even in the presence of forward and backward privacy. Once some information is leaked, it cannot be undone.

**Definition 2.5 (Forward privacy)** *An SSE scheme is forward-private if the file insertion leakage is limited to $\mathcal{L}_{Add}(f) = (id(f), |f|, |\{w\}_{w \in f}|)$ for all new files $f$ being added at any time after running* Build.

This definition states that in a forward-private scheme, the server cannot learn anything about a new file $f$, beyond its identifier and size and the number of its keywords, after any number of searches for the keywords in $f$ before the insertion of $f$ [6]. In other words, the server cannot link a new file to an old temporal access pattern if the scheme is forward-private. In dynamic SSE schemes without forward privacy, the addition leakage would also include the set of keywords in the file that was searched for in the past: $Q \cap \{w\}_{w \in f}$.

Now, we define the information leakage of each protocol.

$\mathcal{L}_{Build}$ shows the information leaked during the build phase:

$$\mathcal{L}_{Build}(F) = (N, n, (id(f), |f|)_{f \in F}).$$

The total number of (keyword, file ID) mappings and the number, identifiers and sizes of all files are leaked.

$\mathcal{L}_{Add}$ shows the leakage during adding a new file $f$:

$$\mathcal{L}_{Add}(f) = (id(f), |f|, |\{w\}_{w \in f}|).$$

The file ID and size and the number of unique keywords in the file leak. The is minimal as a result of forward privacy.

$\mathcal{L}_{Srch}$ shows the leakage during search for a keyword $w$:

$$\mathcal{L}_{Srch}(w, t) = \{F_{w,t}, \text{SP}\}.$$

The temporal access pattern and search pattern are leaked.

Note that in $\mathcal{L}_{Srch}$, $F_{w,t}$ is required to answer the search query in a communication-efficient way, and is true for all efficient SSE schemes. The search pattern leaks since the tokens are deterministic, similarly as in all SSE schemes. Moreover, by obtaining the client's files at the outset, the server learns the number, identifiers and sizes of all outsourced files (unless the files are stored in the ORAM). This is true for all schemes that independently encrypt the files and outsource them to the server. Independent encryption is useful for efficiently decrypting the search results. In all SSE schemes employing this strategy, implicitly all operations leak the related file identifiers as well. Even schemes that do not explicitly show file identifier leakage indeed leak identifiers to be able to operate on the file ciphertexts. Thus, all our leakage is minimal across all known efficient dynamic SSE schemes.

Now, we define the security of our DSSE scheme via ideal-real simulation similar to [28].

**Definition 2.6 (Security of SSE scheme)** *Let* DSSE = (Gen, Build, Add, Search) *be an SSE scheme. The following experiments are executed between a stateful adversary $\mathcal{A}$ and a stateful simulator $\mathcal{S}$ using the leakage functions $\mathcal{L}_{Build}$, $\mathcal{L}_{Add}$, and $\mathcal{L}_{Srch}$:*

- ***Ideal**$_{\mathcal{F},\mathcal{S},\mathcal{Z}}(\lambda)$. An environment $\mathcal{Z}$ sends the client a* setup *message, the set of files to be outsourced and the unencrypted index. The client forwards them to the ideal functionality $\mathcal{F}$. The simulator $\mathcal{S}$ is given $\mathcal{L}_{Build}$.*

  *Later, the environment $\mathcal{Z}$ asks the client to run an* Add *or* Search *protocol by providing the required information. For* Add*, it gives a new file $f$ and the set of unique keywords in the file.* Search *is accompanied with a keyword. The client prepares and sends the respective command to the ideal functionality $\mathcal{F}$. $\mathcal{F}$ gives the corresponding leakages to $\mathcal{S}$. In return, $\mathcal{S}$ sends $\mathcal{F}$ either an* abort *or* continue *command. $\mathcal{F}$ sends the client either $\perp$ (abort) or 'Done' for* Add*, or the set of matching file IDs for* Search*. $\mathcal{Z}$ observes the output. Finally, $\mathcal{Z}$ outputs a bit b as the output of experiment.*

- ***Real**$_{\Pi_{\mathcal{F}},\mathcal{A},\mathcal{Z}}(\lambda)$. An environment $\mathcal{Z}$ sends the client a* setup *message together with the set of files to be outsourced and the unencrypted index. The client runs $\text{Gen}(1^\lambda)$ to generate the key $K$ and starts the* Build *protocol with the real-world adversary $\mathcal{A}$.*

  *Later on, the environment $\mathcal{Z}$ provides the required information and asks the client to start an* Add *or* Search *protocol. For* Add*, it gives a new file $f$ and the set of unique keywords in the file.* Search *is accompanied with a keyword. The client runs the requested protocols with the real-world adversary $\mathcal{A}$. The client outputs either $\perp$ (abort) or 'Done' for* Add*, or the set of matching file IDs for* Search*. $\mathcal{Z}$ observes the client's output. Finally, $\mathcal{Z}$ outputs a bit b as output of experiment.*

We say a DSSE scheme ($\Pi_F$) emulates the ideal functionality $\mathcal{F}$ in the semi-honest model if for all PPT real world adversary $\mathcal{A}$, there exists a PPT simulator $\mathcal{S}$ such that for all polynomial-time environments $\mathcal{Z}$, there exists a negligible function $\nu(\lambda)$ on the security parameter $\lambda$ such that:

$$|\Pr[\mathbf{Real}_{\Pi_{\mathcal{F}},\mathcal{A},\mathcal{Z}}(\lambda) = 1] - \Pr[\mathbf{Ideal}_{\mathcal{F},\mathcal{S},\mathcal{Z}}(\lambda) = 1]| \leq \nu(\lambda).$$

## 3 Construction

In our construction, as with other symmetric searchable encryption schemes, the server stores an encrypted index that relates each keyword to the set of identifiers of files sharing the keyword and helps it perform operations requested by the client. The client stores the number of files containing each keyword and the number of searches per keyword (for forward privacy). To execute a query, the client prepares and sends a token to help the server do the job.

### 3.1 Indexes

Our construction uses data structures, divided between the client and server to maintain the encrypted index. This requires the client and server indexes to be synchronized. While this is not a big problem for our single-client model, it would be problematic for multi-client settings. Outsourcing the client index solves the problem. We first assume the client stores the index, and show how to outsource it as well in Section 3.3.

**Server storage** consists of a dictionary DictW of size $O(N) = O(nm)$. DictW is "indexed" by keywords, and relates each keyword to the set of identifiers of files in which the keyword appears. If $f$ is the $i^{th}$ file containing $w$, $id(f)$ is encrypted using the $K_w$ and $i$, and is stored in DictW at an address depending again on $K_w$ and $i$. (This is detailed in Section 3.2, File addition.) Hence, given a key related to $w$ and the number of files containing $w$, the server finds and decrypts all intended file IDs, and returns the corresponding encrypted files.

We require the DictW to store the entries sorted based on their addresses. Since the addresses are random-looking values generated by a hash function, this ensures that entries of each keyword (and as a result, all entries) are stored in random locations in DictW. As we process the files sequentially to build the DictW (and upload it when all files are processed), this ensures there is no leakage about the entries of each file (or keyword). Note that this only helps to conceal information about the files outsourced at the beginning. Later files insertions reveal the number of unique keywords in the files (discussed in Section 3.2, Leakage), and DictW does not help in this regard.

**Client storage** includes two dictionaries: FileCnt that stores the number of files containing each keyword (as in Cash *et al.* [8]), and SearchCnt that contains the number of times a keyword has been searched for, and is used to generate fresh encryption keys upon search (for forward privacy). Both FileCnt and SearchCnt are of size $O(m)$ and are initialized with zeros. Hence, it is reasonable to store them both locally, compared to the $O(nm)$ outsourced index.

## 3.2 Protocols

Our protocols for setting up the index, adding files, and doing searches are given in Figures 2 and 3, and described next.

**Setup.** We generate two random keys: $K_G$ for a PRF $G$ and $K_{\mathsf{SKE}}$ for a CPA-secure encryption scheme SKE (Gen protocol in Figure 2). These two keys constitute the only cryptographic information stored at the client.

**File addition.** To add a new file $f$, we extract the set of keywords in $f$ and insert a new entry into DictW for each keyword $w_i \in f$. First, FileCnt[$w_i$] is incremented to show that a new file containing $w_i$ is inserted (line 7 of protocol Add). This enables the client to generate consistent tokens later. The new value of FileCnt[$w_i$] also shows the sequence number of $f$ among the files containing $w_i$. This value is used as an input to the hash functions to compute the address in DictW where $id(f)$ will be stored, and to mask its content (lines 8-10 of protocol Add).

The client then encrypts the file, and sends the encrypted file along with the generated set of (key, value) pairs, WPairs, to the server. The server adds the encrypted file into the collection $C$ and inserts WPairs into DictW. Though this is an $O(r)$ operation (where $r$ is the number of unique keywords in the file), the server only copies the token data into its own index.

An important fact about file insertion is that the server does not know any valid (in-use) keyword-related key when a file in being inserted. Hence, it cannot check if this new file contains any keywords from previous searches.

**Building the database.** To initialize the database, the client processes each file, $f_j \in F = \{f_1, f_2, ..., f_n\}$, as described in the Add protocol (but without sending anything to the server), accumulates the results, and uploads the final results to the server altogether. Hence, we present Build as a set of Add protocols, each processing a file $f_j \in F$. This modularity is useful for presentation, but it is important that the actual building is not done in a way that allows the server to observe individual file additions; instead, all files are added to the index by the client and uploaded as a batch. Moreover, the entries of DictW are located according to order of their addresses. This is important for security as otherwise, the server would learn information about individual files/keywords.

For each file $f_j \in F$, we add the encrypted file $c_j$ into the collection $C$ (line 3 of protocol Build), compute WPairs$_j$ and accumulate them into WPairsAll (line 4 of protocol Build). Finally, we send WPairsAll and $C$ to the server that keeps the collection of encrypted files $C$ and stores WPairsAll in DictW.

**Search.** To search for a keyword $w$, the client generates the respective key $K_w = G(K_G, w||\mathsf{SearchCnt}[w])$ and sends it to the server along with the number of files containing $w$: $n_w = \mathsf{FileCnt}[w]$ (lines 1-2 of protocol Search), in a search token $(K_w, n_w)$. The server computes the addresses of DictW entries to be accessed as $h(K_w, i||0)$ and the respective values for unmasking the entries as $h(K_w, i||1)$, for $1 \leq i \leq n_w$. Once the file identifiers are found, the server returns the respective files to the client and deletes[1] the referenced DictW entries (lines 3-7).

---
[1] The server does not need to actually delete the accessed entries, but they have already been revealed and contain no new information. As a fresh key is generated for each search, those entries will not be accessed again.

> Let $G:\{0,1\}^\lambda \times \{0,1\}^* \to \{0,1\}^*$ be a PRF, SKE=(Gen,Enc,Dec) a CPA-secure private-key encryption scheme, and $h: \{0,1\}^* \to \{0,1\}^\lambda$ be hash functions modeled as random oracles, where $\lambda$ is the security parameter.
>
> $(sk)() \leftarrow \text{Gen}(1^\lambda)(1^\lambda)$ :
> 1: $K_G \leftarrow \text{GenPRF}(1^\lambda)$ ▷ For the PRF.
> 2: $K_{SKE} \leftarrow \text{SKE.Gen}(1^\lambda)$
> 3: return $sk = (K_G, K_{SKE})$
>
> $(\mathfrak{I}_c)(\mathfrak{I}_s, C) \leftarrow \text{Build}(sk, F)()$ :
> 1: WPairsAll = {}
> 2: **for all** $f_j \in F$ **do**
> 3:     Run Add to generate $c_j$ and WPairs$_j$. ▷ Without uploading the results.
> 4:     $C = C \cup c_j$ ▷ Add encrypted files to the collection.
> 5:     WPairsAll = WPairsAll $\cup$ WPairs$_j$
> 6: Send $C$ and WPairsAll to the server.
> 7: The server keeps $C$ and stores WPairsAll in DictW.
> 8: $\mathfrak{I}_c = \{\text{SearchCnt}, \text{FileCnt}\}$ and $\mathfrak{I}_S = \{\text{DictW}\}$
>
> $(\mathfrak{I}'_c)(\mathfrak{I}'_s, C') \leftarrow \text{Add}(sk, f, \mathfrak{I}_c)(\mathfrak{I}_s, C)$ :
> 1: WPairs = {}
> 2: **for all** $w_i \in f$ **do**
> 3:     **if** FileCnt$[w_i]$ *is NULL* **then**
> 4:         FileCnt$[w_i] = 0$
> 5:     **if** SearchCnt$[w_i]$ *is NULL* **then**
> 6:         SearchCnt$[w_i] = 0$
> 7:     FileCnt$[w_i]$++ ▷ One more file contains $w_i$.
> 8:     $K_{w_i} = G(K_G, w_i || \text{SearchCnt}[w_i])$
> 9:     $\text{addr}_{w_i} = h(K_{w_i}, \text{FileCnt}[w_i] || 0)$
> 10:     $\text{val}_{w_i} = id(f) \oplus h(K_{w_i}, \text{FileCnt}[w_i] || 1)$
> 11:     WPairs = WPairs $\cup \{(\text{addr}_{w_i}, \text{val}_{w_i})\}$
> 12: $c \leftarrow \text{SKE.Enc}(K_{SKE}, f)$
> 13: Send $c$ and WPairs to the server.
> 14: The server adds $c$ into $C$ and WPairs into DictW.

Figure 2: Protocols for Building and Updating the Database

The client receives the files, increments the number of times $w$ is searched for as SearchCnt$[w]$++, generates a fresh key $K'_w = G(K_G, w, \text{SearchCnt}[w])$ (lines 8-10), re-encrypts the file identifiers with this new key, and sends them back to the server (lines 12-18). These entries are stored at different locations in DictW than the previous ones because their addresses depend on the new key $K'_w$, which includes SearchCnt$[w]$.

The server receives the entries encrypted under a different key and learns that it is the same set of entries accessed recently. Though we use different keys for a keyword $w$ on separate searches, our construction does not eliminate search pattern leakage. However, if a new file $f$ containing $w$ is added after even multiple searches for $w$, the server cannot realize that $f$ contains $w$; satisfying forward privacy. This is because the addresses are pseudo-randomly generated, and without knowing the key, the server cannot infer anything from them.

**Leakage.** Searchable encryption schemes face tradeoffs between performance and leakage, and our scheme reveals some information to the server to enable efficiency and scalability. In Section 4, we provide a formal proof that the proposed scheme satisfies the forward privacy requirements. We now informally discuss leakages of our construction.

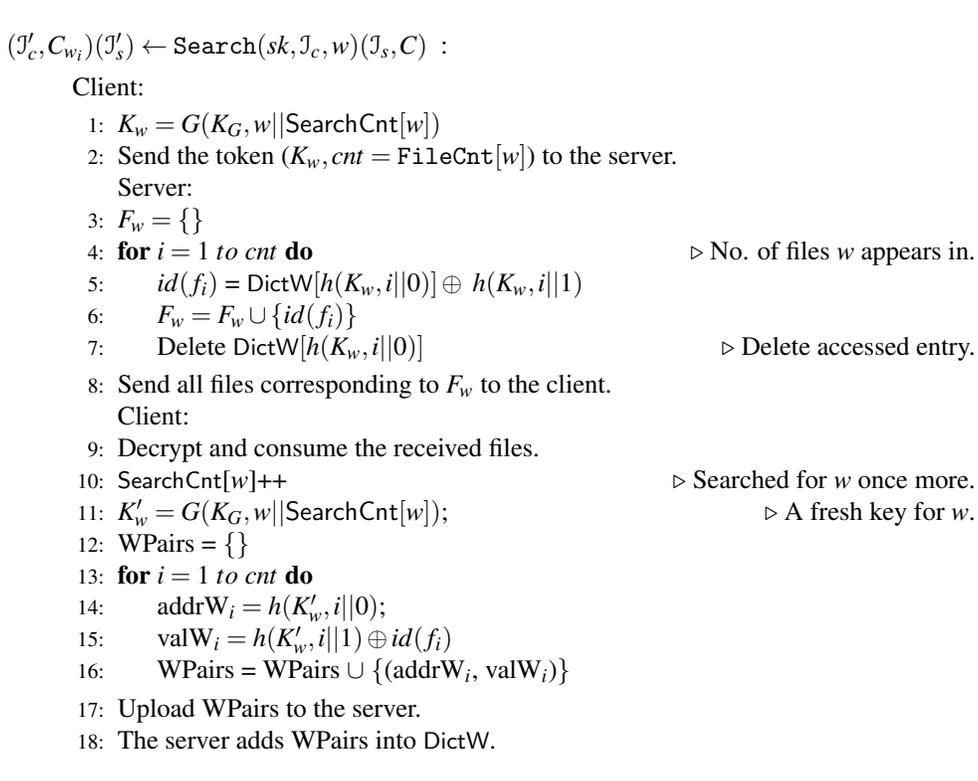

```
(𝔍'_c, C_{w_i})(𝔍'_s) ← Search(sk, 𝔍_c, w)(𝔍_s, C) :
    Client:
     1: K_w = G(K_G, w||SearchCnt[w])
     2: Send the token (K_w, cnt = FileCnt[w]) to the server.
    Server:
     3: F_w = {}
     4: for i = 1 to cnt do                                    ▷ No. of files w appears in.
     5:     id(f_i) = DictW[h(K_w, i||0)] ⊕ h(K_w, i||1)
     6:     F_w = F_w ∪ {id(f_i)}
     7:     Delete DictW[h(K_w, i||0)]                          ▷ Delete accessed entry.
     8: Send all files corresponding to F_w to the client.
    Client:
     9: Decrypt and consume the received files.
    10: SearchCnt[w]++                                          ▷ Searched for w once more.
    11: K'_w = G(K_G, w||SearchCnt[w]);                         ▷ A fresh key for w.
    12: WPairs = {}
    13: for i = 1 to cnt do
    14:     addrW_i = h(K'_w, i||0);
    15:     valW_i = h(K'_w, i||1) ⊕ id(f_i)
    16:     WPairs = WPairs ∪ {(addrW_i, valW_i)}
    17: Upload WPairs to the server.
    18: The server adds WPairs into DictW.
```

Figure 3: Protocol for Search

During the Build protocol, all existing files are processed and encrypted, and the client and server indexes are generated. By uploading the encrypted files, the server learns the number and sizes of all outsourced files. In addition, the size of DictW reveals the number of all keyword-file ID mappings. Since the entries of DictW are ordered by the entries' addresses (which are outputs of a cryptographic hash function), DictW does not leak anything that can be used to infer information about individual keyword/file mappings. Thus, this leakage is the minimum required for an efficient SSE scheme, satisfying the requirements for $\mathcal{L}_{Build}$ in Definition 2.5.

Upon inserting a new file, the file is encrypted, the corresponding set of encrypted DictW entries is generated and uploaded. In addition to the file size, the server learns the number of unique searchable keywords in the file. (The server also learns the time of insertion of each new file, but we consider this outside the scope of the searchable encryption protocol.) In schemes without forward privacy [19, 18, 8, 25], the server learns which previously-searched keywords appear in this new file. Schemes that store the set of identifiers of files sharing each keyword together [25] leak information about the common keywords in the files inserted even before any search takes place. Other schemes supporting forward privacy (including Sophos [6] and Stefanov *et al.*'s scheme [28]) leak the same information as we do. The leakage of our scheme when a file is added is minimal among the known SSE schemes, and is limited to $\mathcal{L}_{Add}$.

A search operation reveals the list of files sharing the keyword under query, which is required for correct responses in SSE protocols. As the queries of different searches for the same keyword vary, it seems that we prevent search pattern leakage. However, the server can link two different queries for the same keyword via the DictW entries that are re-encrypted. Therefore, our search leakage includes the search pattern leakage and the list of intended files, and is limited to $\mathcal{L}_{Search}$. This is also minimal for efficient SSE schemes as pointed to by Cash *et al.* [8].

## 3.3 Outsourcing the Local Index

In the scheme presented, the client stores two local dictionaries that are used for token generation. They are both of size $O(m)$, while the server index is of size $O(nm)$. As an example, with 1M keywords, 4-byte integers, and assuming the average size of the keywords is 10 bytes, the client index would require $1M \times (10 + 4 + 10 + 4) \approx 28MB$ storage. This is manageable for reasonable clients, including any recent smartphone. However, maintaining the client state requires synchronization between the client and server indexes. This would prevent extensions to support multiple clients sharing the server. Hence, we want to outsource this small index as well. Then, the client stores only the keys (a small constant size) for generating the keyword-related keys and encrypting/decrypting files, and no synchronization is necessary at the client.

There are different ways to eliminate the client index, but it should be done cautiously to avoid compromising forward privacy. Note that applying the approaches from Kamara *et al.* [19] and Naveed *et al.* [25] would not preserve forward privacy. Since they store the list of file IDs of each keyword in a separate list, they need to access the corresponding lists on each file insertion to append or add the new file ID. The server learns to which existing lists the new file ID is added, something we need to conceal for forward privacy.

One way to do this would be to use an ORAM. Outsourcing the client index through the ORAM, similar to Sanjam *et al.* [14], is completely consistent with the ORAM definition as it uses only one block per access. This is different from outsourcing the server index using an ORAM, where an operation needs accessing multiple blocks in most cases, which leaks at least the number of accessed blocks. Moreover, ORAM is too expensive to store the whole index, but practical for storing just the SearchCnt and FileCnt.

To search for a keyword $w$ in this setting, the client first reads SearchCnt$[w]$ and FileCnt$[w]$ to prepare and send the respective token to the server. Then, it increments the search counter and updates the ORAM accordingly. (With efficient ORAM implementations, the read and update operations can be combined into one ORAM access, as done by Zahur *et al.* [30].) Finally, the search is done as already described in our construction. For inserting a new file $f$, the client reads the counters FileCnt$[w_i]$ of all keywords $w_i \in f$ through the ORAM (needed for forward privacy), increments them, and prepares and sends the insertion token to the server who updates its index DictW, and the file collection, as described. Finally, it stores the updated counters FileCnt$[w_i]$ inside the ORAM.

## 4 Security Analysis

Our goal is to prove that the proposed scheme provides forward privacy, as defined in Section 2.3, using the leakage functions also defined in that section. We give the proof for the scheme presented in Section 3 assuming the client keeps the local index for simplicity. It would be straightforward to extend it to cover outsourcing the client index as well, but not included here.

**Theorem 4.1** *If* SKE *is a CPA-secure symmetric-key encryption scheme, G is a pseudorandom function, and h is a hash function, our dynamic SSE construction in Figures 2 and 3 is secure based on Definition 2.6, in the random oracle model.*

**Proof 4.1** *To show that the real game is indistinguishable from the ideal game by any PPT distinguisher, we construct a PPT simulator $\mathcal{S}$ who uses the information provided by leakage functions to simulate the client behavior in a way that is indistinguishable form a real client. $\mathcal{S}$ builds simulated versions of both the encrypted server index $\mathcal{I}_S$ and the collection of encrypted files C. Since both $\mathcal{I}_S$ and C are encrypted, even though their actual contents are not known by the simulator, they can be created as random values chosen from a uniform distribution over the range of the encryption scheme or the hash functions used. $\mathcal{S}$ needs information about the number and sizes of the outsourced files, the index size, and the effect of later operations; all are provided by the respective leakage functions.*

*$\mathcal{L}_{Build}$ provides the information required for starting the simulation: the index size N, and the number and sizes of the outsourced files. The simulator creates and fills a dictionary DictW with N randomly-generated values of proper sizes. Since the original contents of DictW are outputs of the random oracle, no PPT distinguisher can distinguish them from the generated random values. All files in the collection C are encrypted. Hence, $\mathcal{S}$ simulates them by encrypting an all-zero strings of size $|f_j|$ for each file $f_j \in F$. The CPA-security of the encryption scheme guarantees that no PPT distinguisher can distinguish this behavior.*

*Simulating the operation tokens (*Add*,* Search*) is more complex. The problem is that these operations affect each other, and $\mathcal{S}$ should keep track of these effects and dependencies among the tokens to keep them consistent, based on information revealed by their respective leakage functions. $\mathcal{S}$ keeps a local copy of DictW, and updates it according* to the information provided by the leakage functions. This local copy is utilized during token generation, and helps generate consistent tokens. Let us illustrate how $\mathcal{S}$ adaptively simulates the encrypted files, indexes, and tokens.

**Initialization.** *The leakage function $\mathcal{L}_{Build}(F)=(N,n,(id(f_j),|f_j|)_{f_j \in F})$ reveals the number and sizes of the existing files and the number of keyword to file ID mappings, N. The simulator takes the following steps:*

```
1: K ← SKE.Gen(1^λ)
2: Generate N pairs (a_i, v_i) randomly and store them in a dictionary DictW. Both a_i and v_i are of length l.
3: Simulate encrypted files as {c_j ← SKE.Enc(K, 0^{|f_j|})}_{f_j ∈ F}.
4: Create a dictionary WKeys to store the last key assigned to each keyword. It can be resized if needed.
5: Create another dictionary WOracle to answer the random oracle queries. WOracle can also be resized over the time.
```

**Simulating the insertion token.** *The simulator uses the information in leakage $\mathcal{L}_{Add}(f) = (id(f), |f|, n_f = |\{w_i\}_{w_i \in f}|)$ to update her local data structures. $\mathcal{S}$ does the followings:*

```
1: for i = 1 to n_f do
2:     Generate random values a_i and v_i, each of length l.
3:     Add a new pair (id(f)||a_i, v_i) into DictW.
4: c ← SKE.Enc(K, 0^{|f|})
5: Output the insertion token: (id(f), c, {(a_i, v_i)}_{i=1}^{n_f}).
```

*Note that due to the forward privacy, $\mathcal{S}$ does not know which already searched keywords $f$ contains. Therefore, we do not assign any keyword to this file at this time. If it appears in a later search result, the actual keyword will be specified by the search leakage. At that point, we will assign the keyword(s) and program WOracle. This is a local decision that does not affect the server.*

**Simulating the search token.** *The leakage $\mathcal{L}_{Srch}(w,t) = \{F_{w,t}, \mathtt{SP}\}$ specifies the set of IDs of files containing the keyword searched for, w. The simulator performs as in Figure 4.*

**Answering random oracle queries.** *When simulating the operations, $\mathcal{S}$ always programs the random oracle matrix WOracle in a consistent way. WOracle queries take three inputs: the last key $K_w$ assigned to the keyword w, a zero or one indicating the address or mask value, and the sequence number of the file in $F_{w,t}$. WOracle[$K_w$][0][i] stores the address of a DictW cell assigned to the $i^{th}$ file ID in $F_{w,t}$, and WOracle[$K_w$][1][i] contains a values used to unmask the value stored at DictW[WOracle[$K_w$][0][i]].*

*Thus, all operations performed by the simulator are polynomial-time operations. Together with the fact that there will be polynomially-many adversary queries at most, it makes the total running time of our simulator polynomial. Besides, the adversary cannot distinguish the outputs of our simulator from those of a real client unless he breaks encryption or distinguishes the PRF output from random.*

# 5 Experiments

This section reports on our results from experiments on datasets up to the full Wikipedia archive. Our scheme outperforms the best previous SSE scheme with forward privacy, and has performance that is comparable with the best existing schemes that do not provide forward privacy.

## 5.1 Experimental Design

The experiments are designed to evaluate the performance of our scheme under load. We outsource data sets scaling up to the full Wikipedia archive (nearly 4M pages), and perform different search and update operations.

**Implementation**. We implemented a prototype of our scheme using C/C++ with the Crypto++ library for cryptographic operations. Our protocols employ only cryptographic hash functions that are instantiated with 20 byte outputs.

```
 1: n_w = |F_{w,t}|                                    ▷ The number of files returned.
 2: if WKeys[w] is NULL then                            ▷ First search for w.
 3:     WKeys[w] ← {0,1}^λ
 4: K_w = WKeys[w]
 5: for i = 1 to n_w do
 6:     if WOracle[K_w][0][i] is NULL then
 7:         if f_i is added after the build phase then  ▷ I.e., there are pairs (id(f_i)||a_i,v_i) in DictW.
 8:             Select an unused (id(f_i)||a_i,v_i) pair.
 9:         else
10:             Select randomly an unused (a_i,v_i) pair.
11:         WOracle[K_w][0][i] = a_i
12:         WOracle[K_w][1][i] = v_i ⊕ id(f_i)
13:     else
14:         a_i = WOracle[K_w][0][i]
15:         v_i = WOracle[K_w][1][i] ⊕ id(f_i)
16:     Remove the pair (a_i,v_i) or (id(f_i)||a_i,v_i) from DictW.
17: NewPairs = {}                                      ▷ Generate the new pairs to be added into DictW.
18: K_w^{new} ← {0,1}^λ
19: WKeys[w] = K_w^{new}
20: for i = 1 to n_w do
21:     Add a new pair (a_i,v_i) into DictW, where a_i and v_i are both random values of length l.
22:     NewPairs = NewPairs ∪ {(a_i,v_i)}
23:     WOracle[K_w^{new}][0][i] = a_i
24:     WOracle[K_w^{new}][1][i] = v_i ⊕ id(f_i)
25: Output the token as (K_w, n_w, NewPairs).
```

Figure 4: Simulating the search token.

Our dictionaries are implemented as C++ maps, which are represented internally as red-black trees. A C++ map stores the pairs sorted, regardless of their initial order. This is very important for our implementation of DictW, so that the initial dictionary leaks no information through the tree structure. Further, all information about the initial files are stored in DictW and outsourced altogether. This means that the server cannot realize how many keywords does each file contain, or how many files does each keyword appear in; though it learns these information gradually through the searches. However, as the later files are added one-by-one, the server observes the set of corresponding entries per file inserted into DictW.

Note that we do not need to store the whole address space for the (keyword, file ID) mappings (e.g., in an array). We store *only* the existing mappings. So, we can have a large address space with a small dictionary. Therefore, collisions may occur, but the actual probabilities are negligible. Let $N$ be upper bounded by $2^{40}$, and we use a hash function with 160-bit outputs. Thus, the address space is $2^{160}$ and we need to store at most $2^{40}$ entries. According to the birthday paradox, the probability of a collision is $\frac{N^2}{2*|address\ space|} = \frac{2^{80}}{2^{161}} = 2^{-81}$. This makes the collision probability vanishingly low, and no collisions were encountered during our experiments.

Nevertheless, it is easy to handle collisions in our scheme. If a collision occurs during the Build phase, we can simply increment the SearchCnt value for a keyword that causes a collision, and re-generate its corresponding entries in DictW before outsourcing. The client cannot detect collisions that occur when adding a new file, however, we would need to rely on the server reporting such collisions to the client, would could then increment the SearchCnt value.

**Network configuration.** Our prototype is a client/server implementation and all client-server communications go over the network, as they would in a real deployment. Our experiments are done on Amazon EC2 using m4.4xlarge instances (64GB of memory, 16 CPU cores) running Ubuntu 16.04 LTS for both the server and client. To evaluate the costs of our scheme, we perform search operations once on a single core and once on five cores. The multi-core version is referred to as the *parallelized* scheme, and the other as the *single-threaded* scheme.

Table 1: Statistical information about our experiments.

| #Files | #Words | DictW entries |
|---|---|---|
| 100,000 | 808,293 | 11,534,529 |
| 250,000 | 1,568,036 | 29,389,776 |
| 500,000 | 2,664,633 | 62,543,206 |
| 750,000 | 3,419,374 | 94,328,341 |
| 1,000,000 | 4,104,976 | 116,065,612 |
| 1,500,000 | 5,390,162 | 178,084,747 |
| 2,000,000 | 6,435,546 | 232,594,077 |
| 2,500,000 | 7,401,246 | 285,650,444 |
| 3,000,000 | 8,300,018 | 341,258,052 |
| 3,835,792 | 9,801,551 | 447,070,889 |

**Dataset.** We use the Wikipedia archive (12/1/2016) with 3,835,792 files (each corresponding to a Wikipedia article) as our test dataset. We repeat the experiments varying the number of files up to the full dataset, leading to different numbers of keyword to file ID) mappings. Table 1 summarizes the experiments, giving the number of files, the number of unique keywords in all files, and the number of DictW entries.

We select all words with lengths between 4 and 10 from all files. The keyword length has no effect on the search cost as they are all processed similarly. We use the Porter stemming algorithm [26] to reduce the keywords to a common form. All keywords, even the mistyped and misspelled ones, are preserved for searchability. The file sizes do not affect the search cost; they affect the insertion cost since the bigger files are expected to have more keywords.

**Comparisons.** We focus our experimental comparisons on Sophos [6], which is the most efficient previous dynamic SSE scheme supporting forward privacy. The Sophos client and server are run on the same two EC2 instances as our scheme, and they communicate through grpc.

We do not include detailed performance comparisons with Blind Storage [25] and Kamara *et al.*'s scheme [19], since neither of those schemes support forward privacy. We do include rough comparisons based on reported performance numbers, and our results appear to be competitive or even better than these schemes (although without being able to test implementation on the same dataset and experimental setup, such comparisons should be interpreted cautiously). We do not have access to an implementation of Stefanov *et al.*'s scheme [28], another SSE scheme supporting forward privacy. This scheme uses an ORAM-based structure with $O(\log N)$ levels to store and access the keyword to file ID mappings. It performs and update operations with $O(d \log^3 N)$ and $O(r \log^2 N)$ cost, respectively, compared to our scheme's costs of $O(d/p)$ and $O(r/p)$. In both cases, we observe a polylogarithmic improvement over this scheme. Moreover, Stefanov *et al.*'s scheme requires a rebuild after $N$ operations that retrieves the whole index (of size $O(N)$). Our scheme does not need rebuild operations and asymptotically outperforms Stefanov *et al.*'s scheme substantially, so it was not necessary to conduct performance experiments to compare them.

## 5.2 Pre-Computation

The client pre-computation time for both Sophos [6] and our scheme is depicted in Figure 5. The numbers are averages from 3 runs. Except where noted, reported values in the rest of this section are based on the experiment with the full Wikipedia archive of 3.8M files and 447M (keyword, file ID) pairs.

There are three pre-computation steps: (1) generating the dictionary and plain index, (2) encrypting the files, and (3) building the encrypted index. The first step is orthogonal to our work and the second step is done similarly by all schemes. Therefore, we focus on the cost of building the encrypted index which differs across SSE schemes.

Over the range of file sizes (Table 1), the average pre-computation time per entry in our scheme is between 30 and 40$\mu s$. This is ∼60% of the cost for Sophos, whose values were between 60 and 70$\mu s$ in our experiments. This pre-computation only needs to be done once, but has considerable cost. For the full 3.8M files experiment (with 3,835,792 files), it took 5h27m with our scheme compared to 8h48m for Sophos. This is due to the higher cost of the asymmetric cryptography used for Sophos' trapdoor, compared with the inexpensive hash functions used by our scheme.

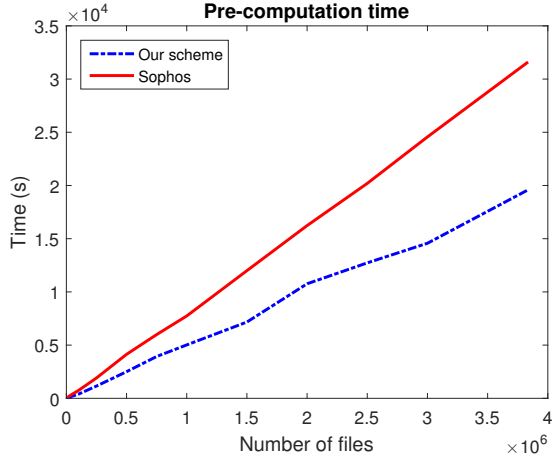

Figure 5: The client pre-computation (setup) time.

Our per entry pre-computation time is comparable to that reported for Kamara *et al.*'s scheme [19]: 35$\mu$s per entry for up to 1.5M total entries. Our per entry time is larger than that for Blind Storage, which starts with 4.13$\mu$s for 5M total entries and reaches 1.58$\mu$s for 20M total entries, since it stores the set of identifiers of files containing a keyword sequentially inside an index file and accesses them altogether. (This is why they cannot have direct file insertion while supporting forward privacy. If two or more files share a keyword, they all need to update the same index file, and the server realizes this fact.)

## 5.3 Search

The client generates a search token that includes the keyword-related key and the number of files in which the keyword appears, and sends it to the server. Token generation is a constant-time operation that does not depend on the number of files or keywords. It takes ∼10$\mu$s in our experiments.

Upon receipt of the token, the server finds and returns all existing files matching the token. Figure 6a shows the search times of queries with different result sizes, on the full 3.8M file dataset. The average per entry search time is ∼10$\mu$s in our single-threaded scheme. Hence, our scheme can perform a search query matching 100,000 files in about one second.

Employing more cores reduces the search time. With five cores running search in parallel, our per entry search time drops to less than 3$\mu s$. As a specific example, using two and three cores the per entry search time with 479,077 files in the result reduces from 9.8$\mu$s to 6.6$\mu$s and 5.2$\mu$s. Figure 6b illustrates how the per entry search time is affected by the number of utilized cores. This shows that our scheme has a very good potential for parallelism (both theoretically and) in practice.

The per-entry search time in Sophos for the same data and queries and in the same settings is ∼15$\mu$s. Despite the fact that Sophos utilizes full multi-threading to parallelize the computations and other optimizations, its per entry search time is slower than our scheme running on a single core. Compared to our parallelized scheme, Sophos's per-entry search time is roughly five times that of our scheme.

Our scheme is also competitive with the best known schemes that do not provide forward privacy. The reported per-entry search time in Kamara *et al.*'s scheme is 7.3$\mu$s. The main source of difference in times is that our scheme needs two hash function evaluations while theirs needs only one. However, their time is more than double the latency of our parallelized scheme. Their scheme is linear in nature and processes the search results one after another, so could not easily take advantage of multiple cores. Blind Storage stores identifiers of all files satisfying each keyword in an index file. Its per entry search time is ∼5$\mu$s, which is half the time required by our single-threaded scheme, but higher than is possible with our scheme using multiple cores.

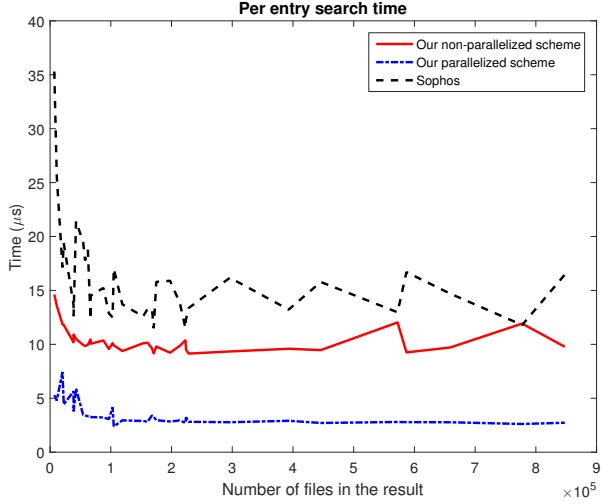
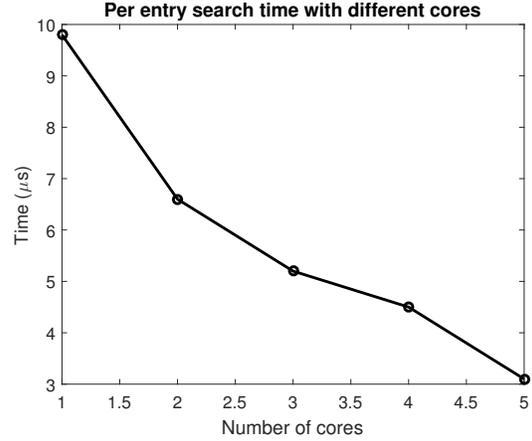

(a) The average per-entry search time.

(b) The per entry search time on different cores, with $n$=3,835,792 total files and $d$=479,077 files in the result.

Figure 6: Search times.

## 5.4 Insertion

To insert a file, the client extracts the keywords in the given file, encrypts the file, prepares and encrypts the new set of (keyword, file ID) pairs, and transfers the results to the server. The server stores the encrypted file and *only* adds the pairs to its indexes. This asymmetry in client and server processing times has already pointed to by Kamara *et al.* [19].

Our file insertion results are in line with what was reported by Kamara *et al.* [19], but a direct comparison is not possible as they use the file (collection) size, while we use the number of unique keywords in the file.

As in the pre-computation phase, processing a (keyword, file ID) pair takes ∼40$\mu$s for the client. Processing a new file with 4133 unique keywords, for example, takes 157 ms for the client and 12 ms for the server, in our experiments. Adding the same file takes 302 ms for the client in Sophos.

Compared to Blind Storage, our scheme still appears to provide better performance. Adding a new file with 2267 unique keywords takes 140 ms in Blind Storage, while a similar operation (adding a file with 2192 unique keywords) takes 81 ms in our experiments.

## 6 Supporting Deletion

Our scheme can be easily extended to support deletion. The Delete protocol is defined formally as:

- $(\mathcal{I}'_c)(\mathcal{I}'_s, C') \leftarrow \texttt{Delete}(sk, id(f), \mathcal{I}_c)(\mathcal{I}_s, C)$: The client uses this protocol to delete a file $id(f)$, given the secret key $sk$ and her current index $\mathcal{I}_c$. It updates the index to $\mathcal{I}'_c$. Similarly, the server takes the index $\mathcal{I}_s$ and file collection $C$ as input, and outputs their updated versions $\mathcal{I}'_s$ and $C'$.

Since DictW is constructed around keywords, to delete a file $f$, the server should examine all DictW entries to find all occurrences of $id(f)$. Hence, we use another dictionary DictF on the server as in Kamara *et al.* [19] that is indexed by file IDs and stores the addresses of DictW entries storing keywords of each file. Now, the server can touch directly the related DictW entries through DictF on each file deletion. This makes updating the DictW on each deletion, efficient. Since DictW and DictF store the same set of information (in different formats), both have the same size N. Similarly, the client needs the number of unique keywords per file to generate consistent deletion tokens. We use another dictionary named WordCnt for this.

One important thing is how to link together the corresponding entries from DictW and DictF. Since a search operation removes some DictW entries, re-encrypts and re-inserts their values into some different entries, the address of those entries stored in the corresponding DictF entries should be updated accordingly. As the contents of DictW entries are encrypted, updating them with the same encryption keys leads to two-time-pad attacks. On the other hand, re-encrypting the data with new encryption information requires storing client-side information per file ID per keyword. To solve this problem, we store *only* the address part of DictW (and DictF) contents in clear. This will reveal nothing more beyond the link between the corresponding DictW and DictF entries. (One can even think of a single table whose entries store contents of the two linked entries together.) Moreover, this will not affect our simulation during the proof.

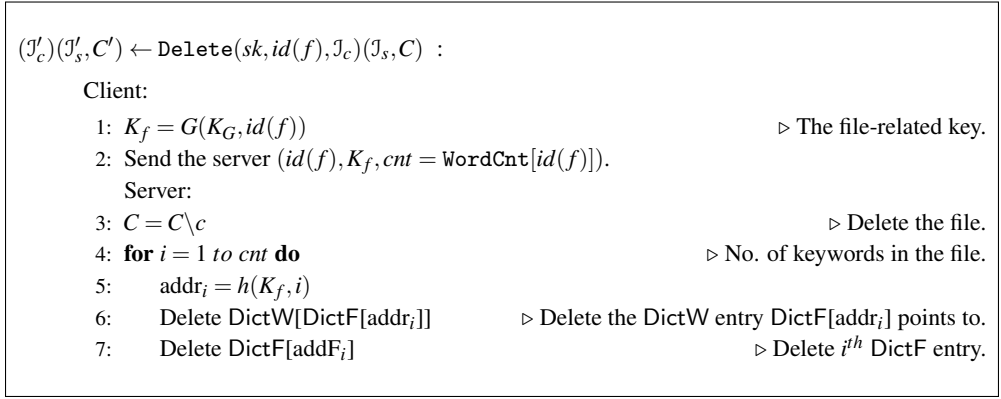

Figure 7: Protocol for Deletion

To delete a file $f$, the server needs a key related to $f$ and the number of keywords it contains. The client provides them inside a deletion token $(K_f = G(K_G, id(f)), n_f = \texttt{WordCnt}[id(f)])$. The server computes $h(K_f, i)$ for $1 \leq i \leq n_f$, to find the addresses of DictF entries related to $f$. Each entry points to a DictW entry. The server locates and removes the respective entries from both DictF and DictW, excluding them from later search results. However, the "location" of the removed DictW entries still belong to the respective keywords. A search operation following the deletion finds such an entry empty and learns that the respective file has already been deleted. Since this search operation removes all DictW entries belonging to the keyword under query, and re-inserts only the valid ones, the subsequent search operations will not see the deleted DictW entries, unless other deletion operations affecting the result come in between. This is an important property of our construction that frees unused memory and does not require heavy rebuild operations, making it very efficient and practical. The related WordCnt entry in the client local index is also removed during deletion. Note that deletion does not update FileCnt. It should be updated during subsequent search operations. The deletion protocol is given in Figure 7[2].

**Leakage.** The server only learns $id(f)$ and the number of unique keywords in $f$. At this time, it cannot link the file to its keywords. However, when a keyword containing the deleted file is searched for (since the deleted DictW cells still belong to the respective keywords), the server realizes that the deleted file contains this keyword. More importantly, the server can learn if some of the already deleted files share the keyword under query, or if a number of deleted files share a keyword searched for in the past; similar to existing works [19, 8, 25]. Backward privacy targets limiting these leakage; we leave backward privacy as future work.

**Performance**. As in search, the client generates and sends the server a deletion token that includes the file-related key and the number of keywords it contains. Token generation is a constant-time operation and does not depend on the number of keywords in the file. The server removes all related pairs matching the token from DictF and DictW, and the related file itself. Figure 8 shows the server times for performing a deletion for both parallelized and non-parallelized schemes, as the number of unique keywords in the file varies (as an indication of the file size). The average per entry deletion time of our non-parallelized scheme is $\sim 17\mu s$ in our experiments. This falls down to $\sim 4\mu s$ in our parallelized scheme, and shows 4X improvement. Deletion is a very fast operation for both the client and server. As an example, deleting a file with 10,460 keywords takes $35ms$ in our parallelized scheme and $147ms$ in our non-parallelized scheme.

Sophos does not support deletion directly. Blind Storage also does not provide direct deletion since they store only the inverted index (i.e., all $F_{w,t}$ sets). They use the *lazy deletion* strategy, i.e., the file indexes will not be updated until the next search operation, and do not report deletion times.

---
[2]Note that the Build, Add, and Search protocols should also be modified to link DictW and DictF entires together.

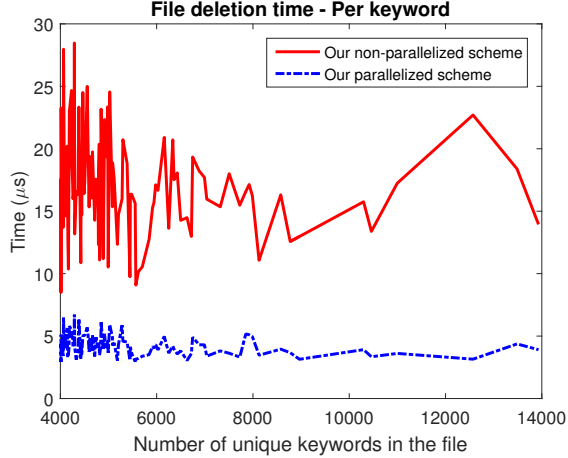

Figure 8: File deletion time by the server.

Our deletion performance is comparable to that of Kamara *et al.*'s scheme. Though they measure the deletion time based on file (collection) size, the main factor affecting the performance is the number of unique keywords in the file. Deleting a large file (e.g., with 10,000 unique keywords) takes ∼140*ms* in our experiments, and Kamara *et al.* report deletion times up to 130*ms* for similar scenarios.

**Effect on asymptotics**. Our scheme, Cash *et al.*'s scheme [8], Blind Storage [25], and Sophos [6] are all asymptotically optimal up to deletions. In Cash *et al.*'s scheme [8] and Sophos [6], the search cost is additionally affected by all file deletions affecting the queried keyword since the beginning ($n_{ad}$). In our scheme and Blind Storage [25], while deletions on the searched keyword affect search performance as well, this effect is neutralized after each search. Thus, our additional cost is related only to *the deletions on the searched keyword since the last search on that keyword* ($n_d$). Obviously, $n_{ad} \geq n_d$. Hence, we achieve even better asymptotic performance, with parallelism and forward privacy.

# 7 Discussion

In this section, we discuss other properties of our scheme.

**Eliminating Random Oracle Assumption.** Our construction is ready to be deployed in the standard model with small modifications: The hash function is replaced by a proper PRF. Then, instead of sending the key and asking the server to compute all respective hash values for search and deletion operations, the client computes and sends all PRF outputs to the server, similar to file addition. The server, given the required PRF outputs, decrypts the requested cells and acts according to the requested operation. Our construction in the standard model inherits and preserves all properties of the random oracle model counterpart. To the best of our knowledge, this is the most efficient SSE construction in the standard model with forward privacy.

Another important advantage of this construction is that the server is not expected to even evaluate hash functions anymore. It only XORs the received values with those in the specified cells to extract the identifiers of files that constitute the answer. This means the server is no longer performing even simple cryptographic operations, and renders our scheme to be deployable in almost all existing cloud environments.

Regarding efficiency, the client's computation and token size for search is increased from $O(1)$ to $O(d)$. The server computation is still $O(d)$, without any hash function evaluations. File insertion continues working in the same manner, i.e., with $O(r)$ client and server computation, and communication. If supported, deletion asymtotics will be increased to $O(r)$, similar to search.

**Parallelism..** Our scheme is ready to benefit from parallelism. It evaluates $O(d)$ and $O(r)$ hash functions for search and update, respectively. Each hash function evaluation is independent, and takes the respective key *k* and a sequence number *s* as input: $h(k,s)$. Therefore, hash function evaluations can all be done in parallel. This allows the service provider to distribute the load on *p* available processors, achieving $O(d/p)$ and $O(r/p)$ search and update costs, respectively. The most efficient known schemes support search with $O(d)$ [19, 6] or $O((d\log n)/p)$ [18] cost and update with $O(r)$ [19, 6] or $O((m\log n)/p)$ [18] cost. Further replication, distribution, and load balancing mechanisms can be employed to improve performance and data availability [29, 27, 2].

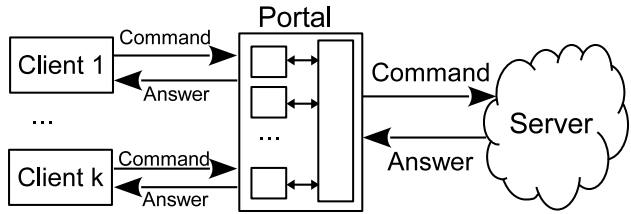

Figure 9: The portal serves all clients.

**Batch update.**. Our scheme in its current state supports batch update (i.e., file insertion). In contrast to the schemes that store index data in a 'sequential' way [19, 25, 28, 6], our scheme stores index data, even for the same file or keyword, in random locations. Therefore, the client can send the updates corresponding to a number of files, without any order, and ask the server to add them all into the index. This also helps to reduce the leakage of adding files one-by-one, since the server does not learn how many keywords each file contains.

**Organizational portal.**. When the number of (keyword, file ID) mappings increases, our scheme requires a larger storage at the client side (if it is not outsourced). In an organization outsourcing a huge number of files, it is not reasonable to replicate the same set of local information over all the clients. Using a *portal server* solves the problem. It is a local (and hence a trusted) entity that stores the same information as a regular client, and serves all clients of the organization. The portal receives the clients' requests, prepares them according to the scheme in use, and sends the resultant command to the server. On receipt the server's answer, relays it to the respective client. In addition, using the local buffer on the portal improve performance of the whole scheme. This is shown in Figure 9.

## 8 Related Work

Several different schemes have been proposed with the general goal of enabling data to be outsourced, while providing some kind of search functionality to data clients. Here, we review those most relevant to our work.

**Oblivious RAM**. Oblivious RAM (ORAM) [15] supports access to an outsourced memory while hiding the access pattern. Different variants of ORAM have been used to minimize the SSE leakages [16, 28, 5, 14]. However, it does not prevent the access pattern leakage [23] since the server needs to learn the list of files to retrieve. To avoid search pattern leakage and provide forward privacy, as in Stefanov *et al.*'s design [28], the server cost scales sublinearly with the total number of (keyword, file ID) mappings (*N*), not just size of the result (*d*).

**Static schemes**. When the outsourced data is intended only for archiving, no update mechanisms are needed. The constructions proposed by Chang and Mitzenmacher [10] support static outsourcing with $O(n)$ search time. Curtmola *et al.* [12] defined CKA2-security for SSE, and proposed adaptively and non-adaptively secure schemes under this definition, with optimal search time, linear in the size of the response. Chase and Kamara [11] gave constructions operating on matrices, labeled data, and graphs. Cash *et al.* [9] support Boolean search.

**Dynamic schemes**. A dynamic SSE scheme provides operations to update encrypted data. Update operations leak more information about the outsourced data. For instance, adding a new file containing a keyword *w* after searching for *w*, reveals to the server that this new file also contains *w* [10, 28]. Table 2 summarizes dynamic SSE schemes.

Kamara *et al.* [19] extended the construction of Curtmola *et al.* [12] to provide a dynamic SSE scheme. They gave a security definition that is adaptively secure against chosen-keyword attacks (CKA2), and presented the first dynamic CKA2-secure construction with optimal search time.

Table 2: A comparison of dynamic SSE schemes.

| Scheme | Client storage | Server storage | Search cost | Update cost | Parallelism | Forward privacy |
|---|---|---|---|---|---|---|
| Kamara *et al.* [19] | $O(1)$ | $O(N)$ | $O(d)$ | $O(r)$ | ✗ | ✗ |
| Parallel SSE [18] | $O(1)$ | $O(mn)$ | $O((d\log n)/p)$ | $O((m\log n)/p)$ | ✓ | ✗ |
| Blind storage [25] | $O(1)$ | $O(N)$ | $O((d+n_d)/p)$ | $O(r/p)$ | ✓ | ✗ |
| Cash *et al.* [8] | $O(m)$ | $O(N)$ | $O((d+n_{ad})/p)$ | $O(r/p)$ | ✓ | ✗ |
| Practical SSE [28] | $O(\sqrt{N})$ | $O(N)$ | $O(d\log^3 N)$ | $O(r\log^2 N)$ | ✗ | ✓ |
| Sophos [6] | $O(m)$ | $O(N)$ | $O(d+n_{ad})$ | $O(r)$ | ✗ | ✓ |
| Ours | $O(m+n)$ | $O(N)$ | $O((d+n_d)/p)$ | $O(r/p)$ | ✓ | ✓ |

$n$ and $m$ denote the total number of files and keywords, respectively. $d$ is the number of files containing a keyword, and $r$ is the number of unique keywords in a file. The number of processors and (keyword, file ID) mappings is $p$ and $N$, respectively. $n_{ad}$ and $n_d$ show the number of times a keyword has been affected by file deletions since beginning and since the last search for the same keyword, respectively ($n_{ad} \geq n_d$).

Kamara and Papamanthou [18] used a red-black tree over a (static) dictionary for building a parallel and dynamic SSE scheme. With $p$ processors running in parallel, it achieves $O((d\log n)/p)$ search and $O((m\log n)/p)$ update cost.

Naveed *et al.* [25] proposed a dynamic SSE scheme using Blind Storage. It encrypts and stores the set of file IDs containing each keyword in a separate index file outsourced through the Blind Storage. While they achieve asymptotic performance and the scheme can be parallelized, they do not offer forward privacy. Indeed, if two files having a common keyword is added, this fact leaks to the server in their scheme.

Cash *et al.* [8] proposed interesting dynamic SSE schemes with asymptotic optimal parallel cost (up to deletions). They extended their static schemes to support file insertion and deletion. After outsourcing the initial data in a static scheme, they use a dynamic scheme (similar to ours) to support later file insertions. The scheme does not support forward privacy. They also store the relation between the deleted files and their corresponding keywords to filter out the deleted files from the search results. This information increases over the time and requires periodic rebuilds to cleanup the indexes.

**Forward-private schemes** The dynamic SSE scheme given by Stefanov *et al.* [28] achieves forward privacy. But, the ORAM-based structure requires the server to checks all levels on each search, leading to the search cost $O(d\log^3 N)$ and update cost $O(r\log^2 N)$.

Sophos [6] supports forward privacy using trapdoor permutation chains. It puts the encrypted file IDs of each keyword in a separate chain. The cost of search and file insertion is as ours asymptotically, but the client and server are expected to run asymmetric cryptography operations. Also, the chain requires sequential scan that prohibits parallelism. To support deletion, they employ another instance of their scheme to keep the list of deleted files and require the sever to operate on both list on each search to filter out the deleted files.

# 9 Conclusion

Ensuring forward privacy is an important step to mitigating attacks on SSE. We propose a dynamic SSE scheme that provides forward privacy with better performance than any previous scheme and without needing any asymmetric operations. Our scheme reduces the required server computation, and limits the server role to mostly storage rather than computation. Hence, our scheme can be employed by a broader range of service providers. Moreover, our design can be converted into a scheme in the standard model. Our scheme is also completely parallelizable and achieves asymptotically optimal search and update costs of $O(d/p)$ and $O(r/p)$, respectively, performing competitively with the most efficient known dynamic SSE schemes that do not provide forward privacy. Although forward privacy is an essential property, it is not sufficient for thwarting all possible attacks on SSE schemes. In particular, it does not address other forms of information leakage and our design does not provide backward privacy. Further progress in these areas is needed before SSE schemes can be used in scenarios where information leakage is unacceptable, but showing that it is possible to achieve forward privacy with high efficiency is an encouraging step towards that goal.

# Acknowledgements

This work was partially funded by National Science Foundation awards 1111781, 1652259, 1526950 and 1514261 and gifts from Amazon and Google as well as a NIST award. We acknowledge the support of TÜBİTAK (the Scientific and Technological Research Council of Turkey) under project number 114E487, European Union COST Action IC1306, and the Science Academy BAGEP Distinguished Young Scientist Award.